\documentclass{article}
\usepackage{spconf,amsmath,graphicx}
\usepackage{booktabs}
\usepackage{multirow}
\usepackage{url}
\usepackage{color}

\title{HIGH EFFICIENCY COMPRESSION FOR OBJECT DETECTION}
\name{Hyomin Choi and Ivan V. Baji\'{c}}
\address{School of Engineering Science, Simon Fraser University, Burnaby, BC, Canada}

\begin{document}
%
\maketitle
\begin{abstract}
Image and video compression has traditionally been tailored to human vision. However, modern applications such as visual analytics and surveillance rely on computers ``seeing'' and analyzing the images before (or instead of) humans. For these applications, it is important to adjust compression to computer vision. In this paper we present a bit allocation and rate control strategy that is tailored to object detection. Using the initial convolutional layers of a state-of-the-art object detector, we create an importance map that can guide bit allocation to areas that are important for object detection. The proposed method enables bit rate savings of 7\% or more compared to default HEVC, at the equivalent object detection rate. 
\end{abstract}
\section{INTRODUCTION}
\label{sec:intro}
Human perceptual quality has always been among the main guiding principles of image and video compression. This influence can be seen throughout the history of development of image and video codecs: from perceptually-optimized quantization matrices in JPEG~\cite{JPEG} to the perceptual rate control~\cite{perceptual_adap_lagrangian_rc, perceptual_rc} for High Efficiency Video Coding (HEVC)~\cite{hevc}. However, modern multimedia applications do not have humans as the only users. In many cases, for example surveillance and visual analytics, computers must ``see'' and examine images or video before humans do. Often, the first step of computer vision would be to detect objects, after which higher-level analytics such as activity recognition or anomaly detection can be performed.    

Despite its importance for these applications, image and video coding tailored to computer (as opposed to human) vision has been largely unexplored. Among the few studies to tackle this topic is gradient-preserving quantization~\cite{GPQ}, which attempts to adjust quantiztion in image compression in order to preserve gradient information. The motivation is that gradients are useful features in a number of computer vision problems, so well-preserved gradients will likely improve the accuracy of the vision pipeline. Another recent work~\cite{sift_feature_rc} develops a rate control scheme for H.264/AVC video coding that preserves SIFT~\cite{sift} and SURF~\cite{surf} features, which have also been found useful in many computer vision problems. These studies (\cite{GPQ,sift_feature_rc}) have proposed ways to preserve well-known handcrafted features through the compression process, without focusing on any particular problem. However, the recent trend in computer vision has been away from handcrafted features and towards learnt features, especially the features learnt by deep neural networks (DNNs)~\cite{Goodfellow-et-al-2016} for specific problems.

In this paper we develop a bit allocation and rate control method that improves object detection of a DNN-based state-of-the-art object detector called YOLO9000~\cite{YOLO2}. We utilize the outputs of the initial convolutional layers of this detector to create the  importance map, which is used to guide bit allocation towards regions that are important for object detection. The resulting strategy offers significant bit savings of 7\% or more compared to the default HEVC at the equivalent object detection rate. For the same bitrate, the proposed strategy offers more accurate object detection and classification compared to the default HEVC.  


The paper is organized as follows. Section~\ref{sec:proposal} describes the creation of object importance maps from the outputs of convolutional layers, and presents the related bit allocation and rate control strategies. Section~\ref{sec:exp} presents the experimental results and Section~\ref{sec:conclusion} concludes the paper.

\section{PROPOSED METHODS}
\label{sec:proposal}

\subsection{Background}
In a convolutional neural network, convolutional layers compute cross-correlation between the input and a set of filters~\cite{Goodfellow-et-al-2016}. The cross-correlation is usually followed by max-pooling, which selects the local maximum within each small window of the cross-correlation output. Large values therefore tend to propagate through the network towards the final layers, where they contribute to the final output. It is important to appreciate that filter coefficients are computed during the training process to maximize the performance on a given task. Hence, DNN-based object detectors have filters whose coefficients have been tuned to extract the features relevant to detecting the objects that the network was trained on. And because max-pooling suppresses small outputs, it follows that large outputs are the ones that are relevant for detection.

The input size of the YOLO9000 object detector~\cite{YOLO2} is fixed at $416 \times 416$. If an input image has different resolution, say $W \times H$, the image resolution is first scaled (while keeping the aspect ratio) and centered so that it fits the input. The scaling constants for various layers are $S_{l} = C_{l}/\text{max}\{W, H\}$, where $C_{l}$ is the spatial dimension of layer $l$, so $C_1=416$, $C_2=208$, etc. 

\begin{figure}[t]
    \begin{minipage}[b]{0.1947\linewidth}
    \centering
    \includegraphics[width=\textwidth]{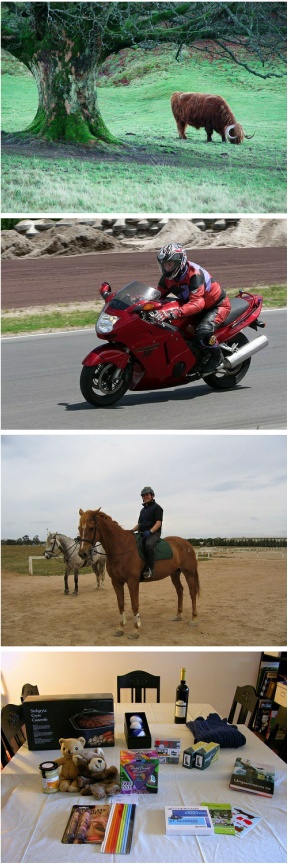}
    \centerline{(a)}\medskip
    \label{fig:a_org_out}
    \end{minipage}
    \begin{minipage}[b]{0.39\linewidth}
    \centering
    \includegraphics[width=\textwidth]{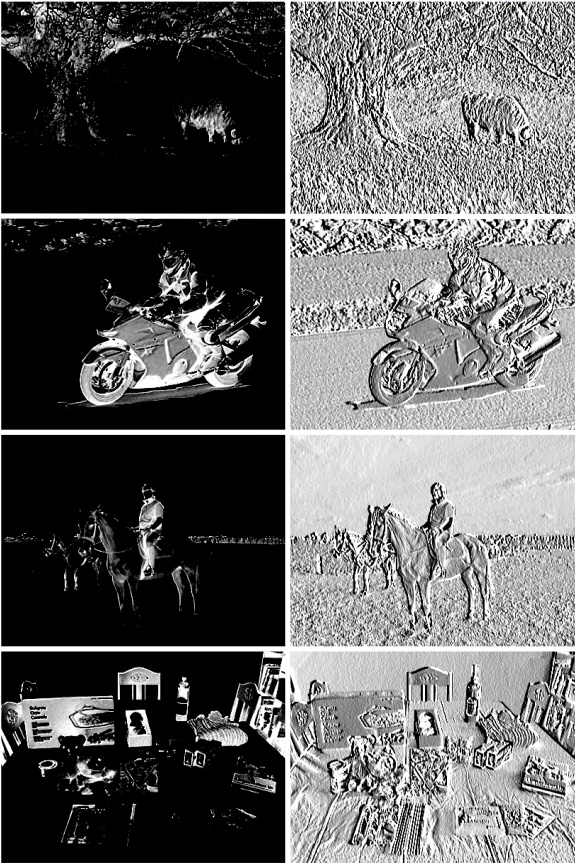}
    \centerline{(b)}\medskip
    \label{fig:b_cnl0_out}
    \end{minipage}
    \begin{minipage}[b]{0.39\linewidth}
    \centering
    \includegraphics[width=\textwidth]{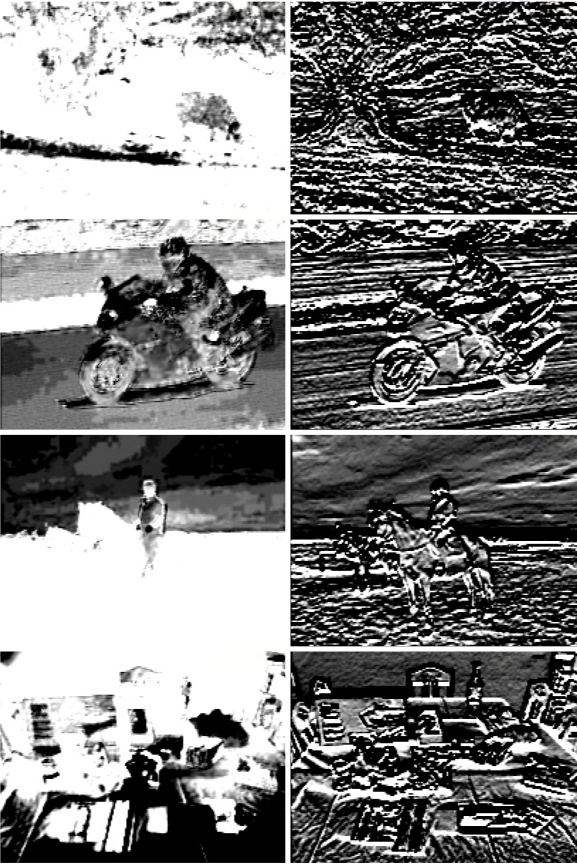}
    \centerline{(c)}\medskip
    \label{fig:c_cnl2_out}
    \end{minipage}
\vspace{-0.7cm}
\caption{Examples of  (a) test images and outputs of selected filters in the (b) first and (c) third  convolutional layers}
\label{fig:convolution_layer_outputs}
\end{figure}

The first convolutional layer employs 32 filters with kernel size $3\times3$, and produces 32 outputs. This is followed by max-pooling over $2 \times 2$ windows. The subsequent convolutional layers operate on the previous layer's outputs. There are total 32 layers in the YOLO9000 architecture. Fig.~\ref{fig:convolution_layer_outputs} shows several input images, and the corresponding outputs of several filters in the first and third convolutional layer. 
The brighter pixels in the output indicate the higher correlation with the associated filter. As seen in this figure, even the early layers in the convolutional network are able to provide some information about the objects, although precise object location and class is not available until upper layers of the network complete their processing. 

\begin{figure}[t]
\centerline{\includegraphics[width=8.5cm]{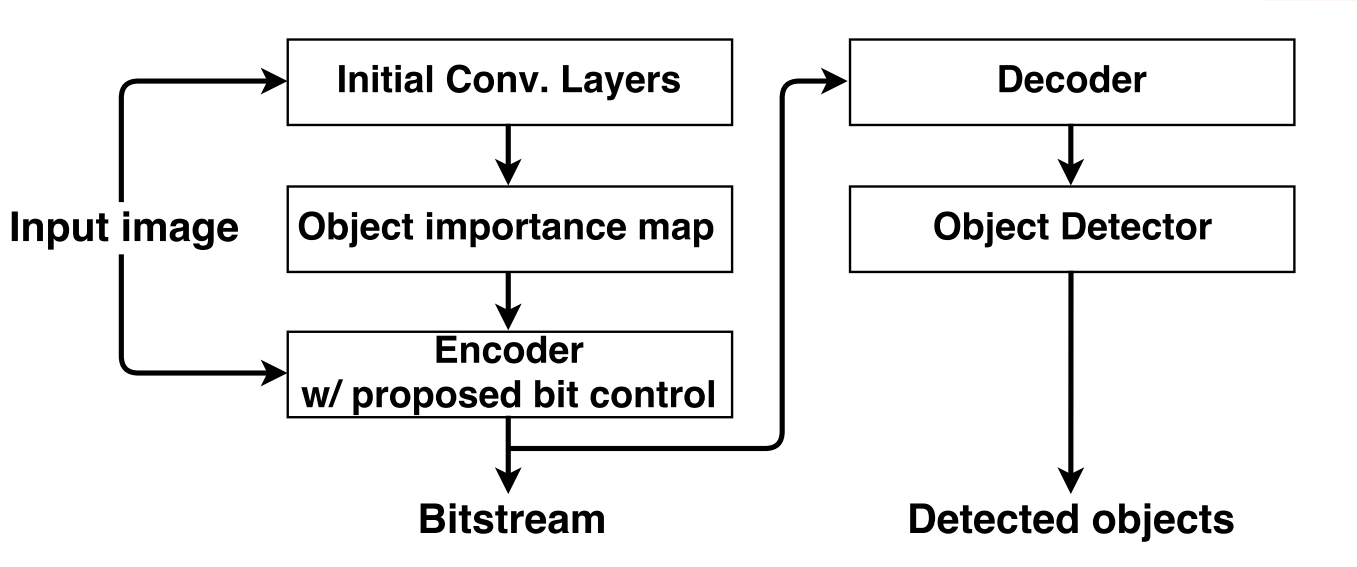}}
\caption{The proposed object detection-friendly compression framework}
\label{fig:flowchart}
\end{figure}

Based on this reasoning, we propose an object detection-friendly compression framework shown in Fig.~\ref{fig:flowchart}. The input image is processed by the initial convolutional layers of the object detector. The filters in each layer can run in parallel, so this process is highly parallelizable. From the resulting filter outputs, we construct an object importance map, which guides bit allocation and rate control in HEVC. The resulting image turns out to be more object detection-friendly, as demonstrated in Section~\ref{ssec:obj_impt_map}.   

%

\subsection{Object importance map}
\label{ssec:obj_impt_map}

The object importance map is meant to indicate how important is each pixel to object detection. The YOLO9000 architecture employs leaky activation, which means that layer outputs can be negative. We first clamp the outputs to the range $[0,1]$ as 
\begin{equation} 
\hat{\psi}_{l}^{\left(n\right)}\left ( x,y \right ) = \max\left \{0, \min \left \{1, \psi_{l}^{\left(n\right)}\left ( x,y \right ) \right \} \right \}
\label{eq:cliping}
\end{equation}
\noindent where $l$ indicates the layer, $x$ and $y$ are spatial coordinates, and $n$ is the filter index in the given layer. Then, all clamped outputs are stacked in a tensor
\begin{equation}
\resizebox{1.0\hsize}{!}{$
v_{l}\left ( x,y \right ) =
\left [  \alpha _{l}^{\left(1\right)}\hat{\psi}_{l}^{\left(1\right)}\left ( x,y \right ), \alpha _{l}^{\left(2\right)}\hat{\psi}_{l}^{\left(2\right)}\left ( x,y \right ), \ldots, \alpha _{l}^{\left(N\right)}\hat{\psi}_{l}^{\left(N\right)}\left ( x,y \right ) \right ]
$}
\label{eq:1d_output_vec}
\end{equation}
\noindent  where $\alpha_{l}^{\left(n\right)}$ is a weight factor for the $n$-th filter in layer $l$. 

\begin{figure}[t]
    \begin{minipage}[b]{0.49\linewidth}
    \centering
    \includegraphics[width=\textwidth]{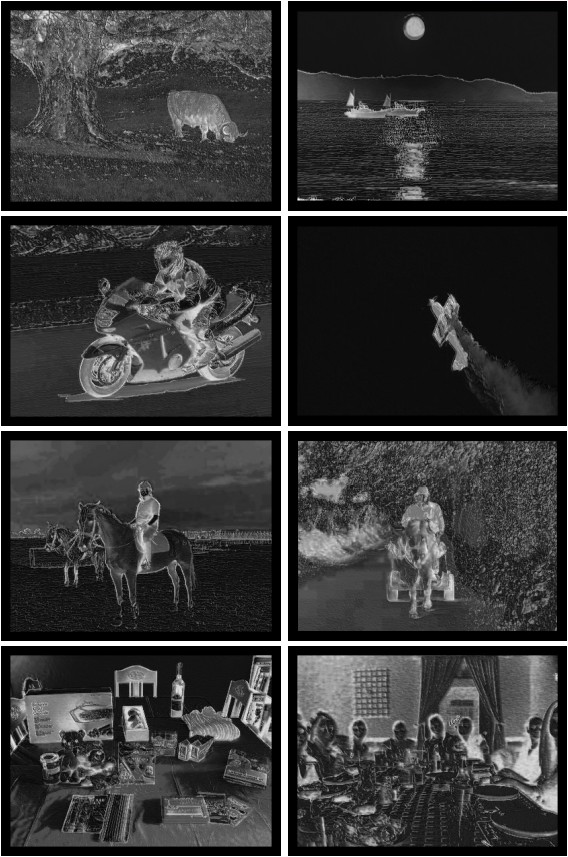}
    \centerline{(a)}\medskip
    \label{fig:syn_cnl0}
    \end{minipage}
    \begin{minipage}[b]{0.49\linewidth}
    \centering
    \includegraphics[width=\textwidth]{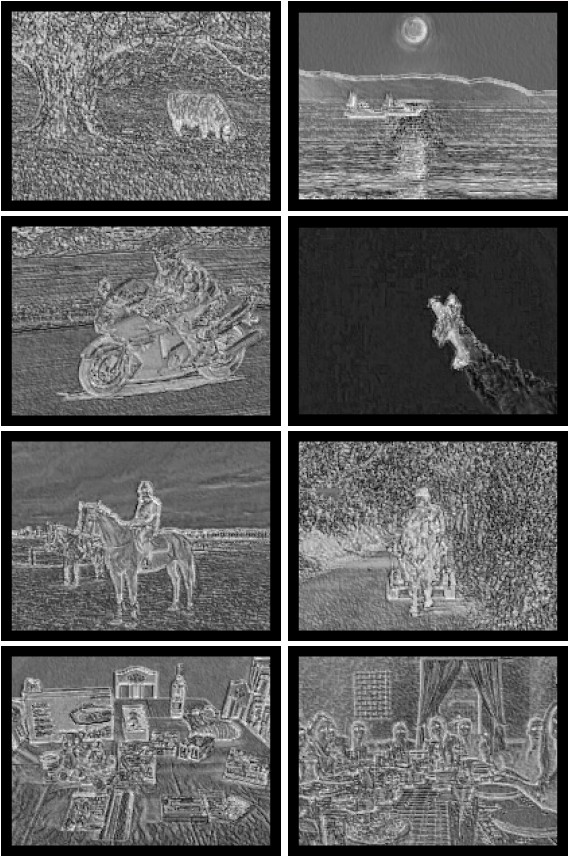}
    \centerline{(b)}\medskip
    \label{fig:syn_cnl2}
    \end{minipage}
\vspace{-0.6cm}
\caption{The object importance maps combined using the outputs of the (a) first and (b) third layer}
\label{fig:object_importance_map}
\end{figure}

The weights are meant to indicate how informative is a particular filter's output for a given input image. Ideally, the filter's output would be high near the objects of interest and low elsewhere. As seen in Fig.~\ref{fig:convolution_layer_outputs}, filters' outputs are not equally informative about the objects in the image. Moreover, a certain filter may be very informative on one image, and not very informative on another image, which means that weights should be adapted from image to image. We experimented with entropy of the filter output as a guide to set weights (lower entropy inducing higher weight), but eventually settled for a simpler approach that gave slightly better results. In particular, we set the weight as 1 minus the average clamped output:
\begin{equation} 
\alpha _{l}^{\left(n\right)} = 1- \frac{1}{{W_{l}\cdot H_{l}}} \sum_{y=0}^{H_{l}-1}\sum_{x=0}^{W_{l}-1}\hat{\psi}_{l}^{\left(n\right)}\left ( x,y \right )
\label{eq:weighting_factor}
\end{equation}
\noindent where $W_l=W/S_l$ and $H_l=H/S_l$ are the width and height of the filter's output on the particular image at level $l$. When the filter produces high responses across the entire image (i.e., it is not very informative), the average is high, so its weight becomes low. If the filter's output is low on average, its weight becomes high. Therefore, $v_{l}(x,y)$ in Eq.~(\ref{eq:1d_output_vec}) will be high only when the filter is informative (high weight) and has a high response at the particular $(x,y)$.
Finally, we take the $\ell^2$ norm of $v_{l}(x,y)$, $O_l(x,y) = \| v_l(x,y) \|_{2}$,  
and then normalize $O_l(x,y)$ by linearly mapping it to the range $[0,1]$ to produce the final importance map $\tilde{O_{l}}(x,y)$. Figure~\ref{fig:object_importance_map} shows several importance maps generated from the first and third layer on different images. 

\subsection{Bit allocation and rate control}
\label{ssec:bit_allocation}

The proposed bit allocation makes use of the object importance map $\tilde{O_{l}}(x,y)$ to decide how to spend bits. 

First, the pixel-wise importance $\tilde{O_{l}}(x,y)$ is converted into block-wise importance $I(i,j)$. The size of the block is the size of the corresponding coding unit scaled by $S_l$. Then $I(i,j)$ is computed by summing $\tilde{O_{l}}(x,y)$ within the corresponding block and dividing by the total sum over the importance map.
From here on, we use $(i,j)$ as the coordinates of the top-left corner of the block. 

We then calculate the initial coarse estimates of the bits per pixel ($bpp_{coarse}$) for each block as 
\begin{equation} 
bpp_{coarse}\left(i, j \right )=\frac{1}{N_{pixels}\left(i, j\right)} \cdot I\left(i, j \right ) \cdot T_{bits}
\label{eq:coarse_bpp}
\end{equation}
\noindent where $N_{pixels}(i,j)$ is the number of pixels in the block whose top-left corner is at $(i,j)$ and $T_{bits}$ represents the number of target bits for the image to be coded.
In Eq.~(\ref{eq:coarse_bpp}), the calculated $bpp_{coarse}(i,j)$ could possibly be zero, which turns out to be harmful in subsequent encoding. In order to refine this coarse estimate, we run the R-$\lambda$ model~\cite{r_lambda, qp_refinement}, which is default rate control model in the HEVC reference software~\cite{HM16.12}. Specifically, by inputting $T_{bits}$ and $bpp_{coarse}(i,j)$ into the R-$\lambda$ model, we compute the slice/picture-level $QP_s$ and the block-level preliminary quantization parameter $QP_p(i,j)$. Note that $QP_p(i,j)$ is bounded by $QP_s\pm2$ by default, however we extend the range to $QP_s\pm3$. Then, the preliminary bits per pixel $bpp_p(i,j)$ is computed by inverting the R-$\lambda$ model with $QP_p(i,j)$ as the input. Finally, $QP_p(i,j)$ is incremented by $1$ if $I\left(i,j\right) = 0$. 

The final block importance is computed as
\begin{equation} 
I_{F}\left(i,j \right) = \frac{bpp_{p}\left(i,j \right)}{\sum\sum bpp_{p}\left(i,j \right)}
\label{eq:fine_importance}
\end{equation}
where the double summation is over all valid $(i,j)$. Using this, we compute the weight for each block as the normalized importance of that block
\begin{equation} 
w\left( i,j \right) = \frac{I_{F}\left(i, j \right )}{\sum\sum I_{F}\left(i,j \right )}
\label{eq:weighting}
\end{equation}
where again the double summation is over all valid $(i,j)$. Then the total target bits for the block can be computed as 
\begin{equation} 
\resizebox{1.0\hsize}{!}{$
T_{bits}^{Blk} \left(i,j \right )= \left \{ L_{bits}(i,j) + \frac{L_{blk}(i,j) \left(L_{bits}(i,j) - L_{bits}^{Est}(i,j)\right )}{SW}\right \}w(i,j)
$}
\label{eq:bit_alloaction}
\end{equation}
\noindent where $L_{bits}(i,j)$ are the remaining bits in the bit budget before coding block $(i,j)$, $L_{blk}(i,j)$ is the number of remaining blocks to be coded, including $(i,j)$, and $SW$ is the sliding window size used to smooth out the bit variation (we use the default $SW = 4$ from the HM16.12~\cite{HM16.12}). 
$L_{bits}^{Est}(i,j)$ is an estimate of the bits that will actually be used for coding the block $(i,j)$ and all subsequent blocks, computed as
\begin{equation} 
L_{bits}^{Est}(i,j) = \sum_{u\geq i} \sum_{v\geq j} bpp_{p}\left(u,v \right )\cdot N_{pixels}\left(u,v \right )
\label{eq:left_bit_est}
\end{equation}
\noindent where the summations are starting at $(i,j)$ and going over all subsequent valid block indices $(u,v)$.



Finally, we estimate the actual QP values $QP_a$ by inputting $T_{bits}^{Blk}(i,j)/N_{pixels}(i,j)$ to the R-$\lambda$ model. The resulting $QP_a$ is bounded by $QP_s\pm2$ by default. However, we shift the bound upward as $[QP_s, QP_s+4]$ if $QP_{p} \geq QP_s+3$. This has the potential to save bits in less important regions.

\section{EXPERIMENTS}
\label{sec:exp}

\begin{figure*}[t]
    \begin{minipage}[b]{0.5\linewidth}
    \centering
    \centerline{\small HM (QP=22) \hspace{1.3cm} HM-RC \hspace{1.7cm} Proposed}
    \includegraphics[width=\textwidth]{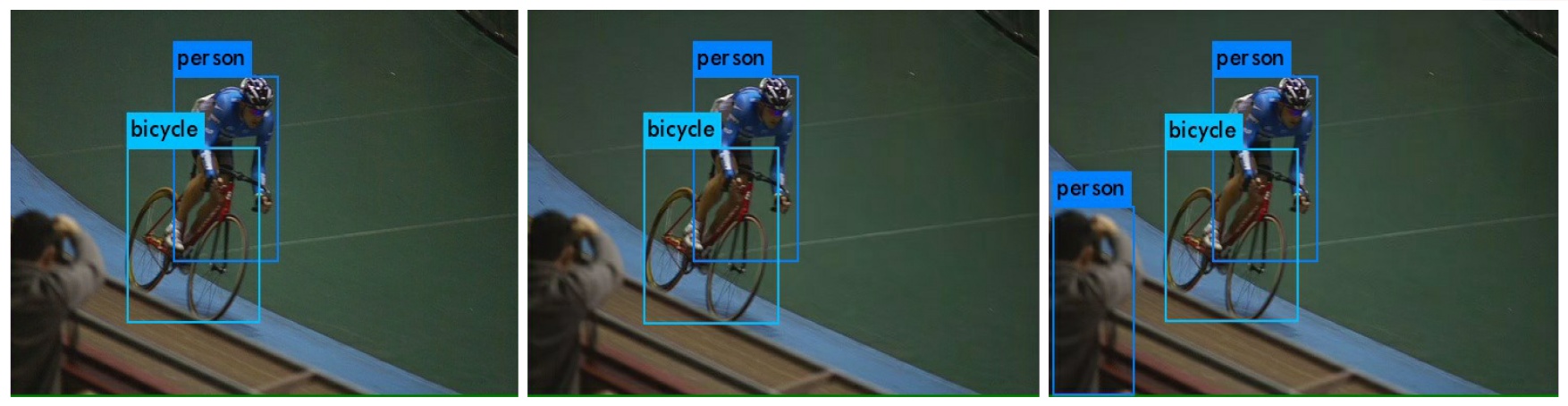}
    \centerline{Increasing correct detections}\medskip
    \label{fig:a_find_new}
    \end{minipage}
    \hspace{-0.18cm}
    \begin{minipage}[b]{0.5\linewidth}
    \centering
    \centerline{\small HM (QP=22) \hspace{1.3cm} HM-RC \hspace{1.7cm} Proposed}
    \includegraphics[width=\textwidth]{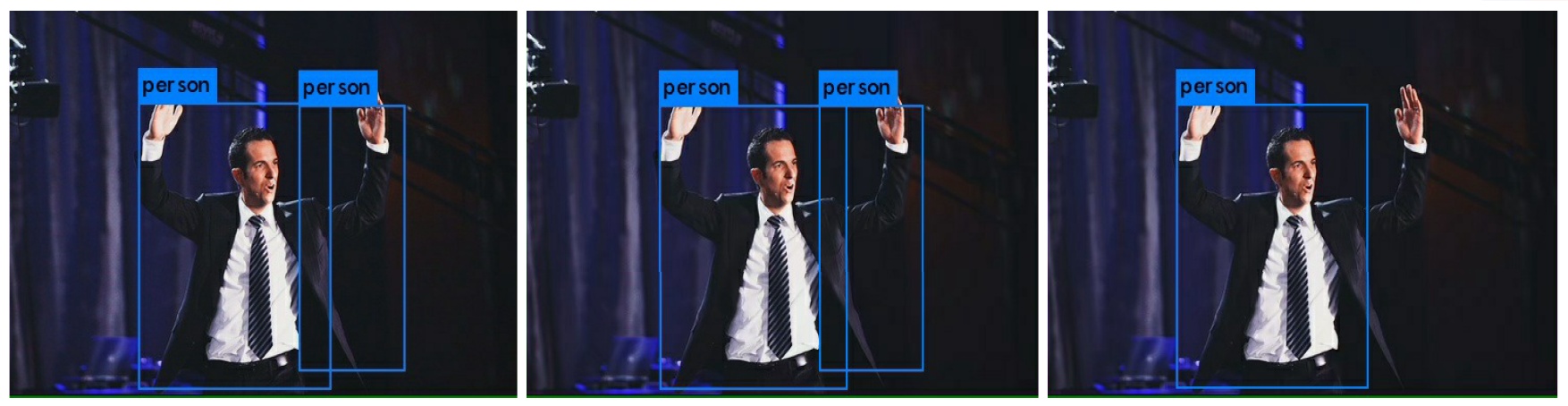}
    \centerline{Decreasing false detections}\medskip
    \label{fig:b_clear}
    \end{minipage}
    \hspace{-0.18cm}
    \begin{minipage}[b]{0.5\linewidth}
    \centering
    \includegraphics[width=\textwidth]{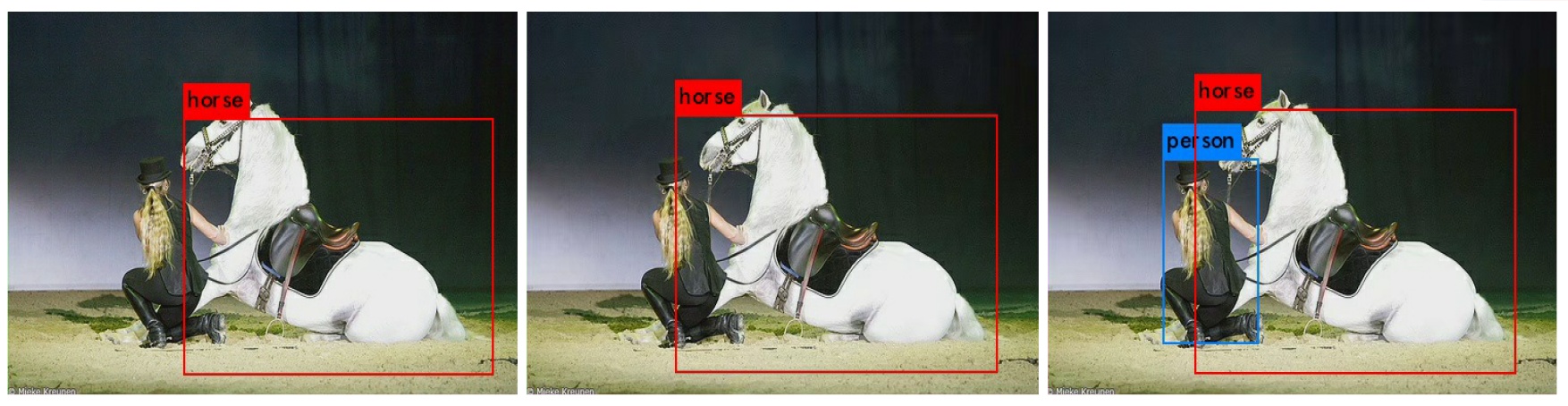}
    \centerline{Increasing correct detections}\medskip
    \label{fig:c_alter_obj}
    \end{minipage}
    \hspace{-0.18cm}
    \begin{minipage}[b]{0.5\linewidth}
    \centering
    \includegraphics[width=\textwidth]{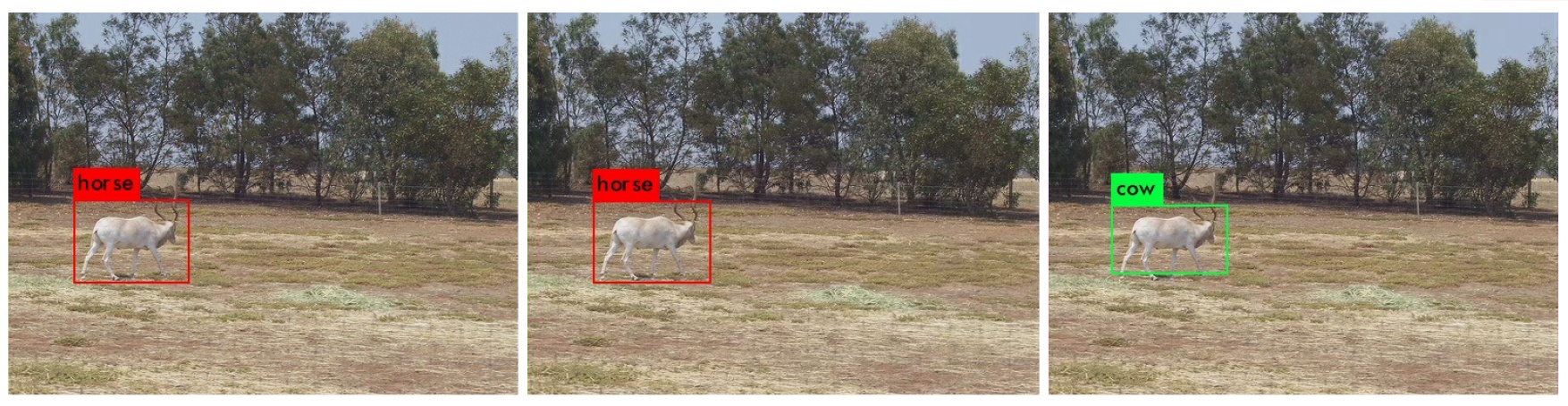}
    \centerline{Alternative object class}\medskip
    \label{fig:d_alter_class}
    \end{minipage}
\vspace{-0.5cm}
\caption{Examples illustrating different object detection/classification produced by YOLO9000 on images encoded by HM with QP=22, HM-RC at the same bitrate, and the proposed method at the same bitrate, with importance maps computed from the third convolutional layer.}
\label{fig:final_outcomes}
\end{figure*}

In this section, we assess the performance of the proposed bit allocation and rate control scheme in terms of its effect on object detection. 
The proposed methods were implemented in HEVC reference software HM16.12~\cite{HM16.12}. The YOLO9000 model in the Darknet framework~\cite{darknet} is used for the object detection performance evaluation. 

Bj{\o}ntegaard Delta (BD)~\cite{bd_br} is a standard measurement method for evaluating compression performance. It compares the average bit rates of two coding methods at the equivalent quality metric. Usually, the quality metric is Peak Signal-to-Noise Ratio (PSNR) and we refer to this measurement as BD bitrate for PSNR (BD-BR-PSNR). However, it is also possible to use quality metrics other than PSNR in the BD analysis. Specifically, since our goal is to compare object detection performance between methods, instead of PSNR we use the standard object detection accuracy metric called mean Average Precision (mAP)~\cite{pascal-voc-2007}. The mAP is in the range $[0,1]$. By computing BD over rate vs. mAP curves, we can obtain the average bit rate saving (or increment) that one compression method would have over another at the equivalent mAP. We call this metric BD bitrate for mAP (BD-BR-mAP).

For testing, we employ the widely used PASCAL VOC 2007 dataset~\cite{pascal-voc-2007},
which has 9963 images out of which 4952 are test images. The images are annotated with 20 different object classes, such as aeroplane, bicycle, bird, and so on. For encoding, 16$\times$16 CTU is adopted and RDOQ tool is off, but other coding parameters follow the common HEVC test conditions~\cite{hevc_ctc} of the Main Still Picture Profile~\cite{HEVC_MSP}. We first encode each test image using the default HM with QP$\in \left \{22, 27, 32, 37 \right \}$. The resulting bits are used as target bits for the default HM rate control (HM-RC) and our proposed method. For the proposed method we construct the importance maps from the outputs of the first, third, and seventh layer, in order to examine the behaviour of the system with different importance maps.  

All encoded images are then decoded and fed to the YOLO9000 object detector. mAP is computed by comparing detector's output with the ground truth. 
Table~\ref{tbl:overall_performance} shows various comparisons among the three tested codecs: HM, HM-RC and proposed. For the rate control accuracy, $\Delta_{bpp}$ is the mean absolute difference (MAD) in bits per pixel (bpp) between the output bits of HM and the two rate control methods (HM-RC and proposed) across all images. HM-RC shows averaged $\Delta_{bpp} = 0.0483$, while our rate control gave smaller deviation in each of the three cases, with importance map computed from the seventh layer being the most accurate.  
In terms of BD-BR-PSNR, both HM-RC and our proposed method have lower performance (positive BD-BR-PSNR) compared to the default HM, since they both deviate from the optimal rate-distortion allocation in order to achieve different objectives. However, our method achieves significant advantage in BD-BR-mAP over both HM and HM-RC, which was the main design objective. In particular, with the importance map computed from the output of the third convolutional layer, 7.32\% bit reduction is achieved over HM, and 8.23\% reduction over HM-RC, at an equivalent mAP. This shows that importance maps can successfully guide bit allocation towards regions that are most relevant for object detection. 
Fig.~\ref{fig:final_outcomes} shows a few examples where YOLO9000 produces different detections on the images encoded by HM, HM-RC, and proposed methods. Although the images look very similar visually, detection on images encoded by the proposed method is the most accurate. This again illustrates that importance maps successfully guide bit allocation towards regions that are most relevant for object detection. 

\begin{table}[b]
\centering
\caption{Performance comparison among HM, HM-RC, and proposed method}
\label{tbl:overall_performance}
\smallskip\noindent
\resizebox{\linewidth}{!}{%

\begin{tabular}{lcccc}
\hline
\multicolumn{1}{c}{\multirow{2}{*}{Test cases}} & \multicolumn{1}{c}{\multirow{2}{*}{$\Delta_{bpp}$}} & \multirow{2}{*}{$\sigma_{bpp}$} & \multirow{2}{*}{BD-BR-PSNR} & \multirow{2}{*}{BD-BR-mAP} \\
\multicolumn{1}{c}{}                         & \multicolumn{1}{c}{}                  &                   &                             &                          \\ \hline
HM vs. HM-RC                                 & 0.0483                                & 0.1187            & 3.08\%                      & 1.67\%                   \\ \hline
HM vs. Ours w/ 1st L.                        & 0.0385                                & 0.1113            & 7.10\%                      & \textbf{-3.90\%}       \\ \hline
HM vs. Ours w/ 3rd L.                        & 0.0372                                & 0.1094            & 7.15\%                      & \textbf{-7.32\%}       \\ \hline
HM vs. Ours w/ 7th L.                        & 0.0232                                & 0.1086            & 6.96\%                      & \textbf{-6.33\%}       \\ \hline
HM-RC vs. Ours w/ 1st L.                     & -                                     & -                 & 3.82\%                      & \textbf{-5.31\%}       \\ \hline
HM-RC vs. Ours w/ 3rd L.                     & -                                     & -                 & 3.87\%                      & \textbf{-8.23\%}       \\ \hline
HM-RC vs. Ours w/ 7th L.                     & -                                     & -                 & 3.68\%                      & \textbf{-7.10\%}       \\ \hline

\end{tabular}}
\end{table}

\section{Conclusion}
\label{sec:conclusion}

We proposed a novel bit allocation and rate control strategy whose goal was to improve object detection after decoding. Using the outputs of the initial convolutional layers of a state-of-the-art object detector, the proposed algorithm successfully achieved efficient bit control and improved object detection performance over the default HEVC implementations. The proposed strategy can be used in
many applications where computers ``see'' and analyze the data before (or instead of) humans.


\bibliographystyle{IEEEbib}
\bibliography{ref}

\end{document}